# Airborne TDMA for High Throughput and Fast Weather Conditions Notification

Hyungjun Jang*, Hongjun Noh, Jaesung Lim

Graduate School of Network Centric Warfare*, Information and Communications,
Ajou University

ajouajou@ajou.ac.kr, nonoboy@ajou.ac.kr, jaslim@ajou.ac.kr

```
San 5, Woncheon-dong, Youngtong-Gu, Suwon city, South Korea
```

**ABSTRACT**

*As air traffic grows significantly, aircraft accidents increase. Many aviation accidents could be prevented if the precise aircraft positions and weather conditions on the aircraft's route were known. Existing studies propose determining the precise aircraft positions via a VHF channel with an air-to-air radio relay system that is based on mobile ad-hoc networks. However, due to the long propagation delay, the existing TDMA MAC schemes underutilize the networks. The existing TDMA MAC sends data and receives ACK in one time slot, which requires two guard times in one time slot. Since aeronautical communications spans a significant distance, the guard time occupies a significantly large portion of the slot. To solve this problem, we propose a piggybacking mechanism ACK. Our proposed MAC has one guard time in one time slot, which enables the transmission of more data. Using this additional data, we can send weather conditions that pertain to the aircraft's current position. Our analysis shows that this proposed MAC performs better than the existing MAC, since it offers better throughput and network utilization. In addition, our weather condition notification model achieves a much lower transmission delay than a HF (high frequency) voice communication.*

## 1. Introduction

Air traffic is expected to increase significantly in the coming decades [1]. Also the increase of UAVs (unmanned aerial vehicle) [15] and small general aviation airplanes accelerates air traffic. As more people fly, the number of flights will steadily increase and the potential for aircraft accidents will increase. In 2009, thirty aircraft accidents occurred, and fatal accidents have occurred since 1950. One of major cause of these accidents is the weather. In fact, thirty percent of commercial aircraft accidents cite weather conditions as a contributing factor [4].

As air traffic increases, aircraft communication will become increasingly important. Many accidents can be avoided with precise aircraft position and weather report information. To increase air transportation safety, we need to study oceanic communication and provide better aircraft position information and weather information. We need a weather condition communication system that is faster and more reliable than HF voice communication.

Providing a better communication system will affect all types of aeronautical communication, including both of its main components. The first main component is the safety critical Air Traffic Service (ATS) and Aeronautical Operational Control (AOC) communication; and the second component is the non-safety critical Aeronautical Administration Communication (AAC) and Aeronautical Passenger Communication (APC). Each of these main areas can be further subdivided. The ATS can be subdivided into Aeronautical Traffic Control (ATC), Flight Information Service (FIS), and Alert. The ATC can be subdivided into Continental Area Communication (CAC) and Oceanic Communication.

In continental areas, the ATC currently uses a narrowband Very High Frequency (VHF) voice system combined with a VHF digital data link such as a VHF Digital Link (VDL) Mode 2,3 [2] or Aircraft Addressing and Reporting System (ACARS). In remote areas and over oceans, High Frequency (HF) and Satellite Communications (SATCOM) voice and data link systems are used [5]. In oceanic and polar areas, a long distance communication system in the HF band is used. However, HF communication has several shortcomings. For example, the HF band depends upon the ionosphere for its sky wave coverage pattern; therefore, while it can provide a multi-path communication beyond the horizon, the operating frequencies must be adjusted according to different weather conditions.

In oceanic and polar areas, a long distance aeronautical communications model has been developed for TDMA MAC. Ho Dac Tu [5] proposed using local mobile ad-hoc networks based on air-to-air links once any two airplanes are within communication distance. These mobile ad-hoc networks are established locally on each aircraft. This TDMA MAC uses only one aeronautical VHF channel with an air-to-air radio relay system based on local mobile ad-hoc networks. However, this existing TDMA MAC is not efficient. Because aeronautical communication is a long distance communication system, therefore, there are considerable propagation delays. In addition, the guard time length is proportional to the communication distance and aeronautical communications based on theoretical evaluation, which is a basic communication distance of 678 km [5]. So, the aeronautical communication time slot needs 3.2 msec of guard time. When an aircraft sends data, it waits until it receives ACK in one time slot. This scheme needs two guard times in one time slot. Thus, this scheme is not efficient in terms of utilization.

To overcome this problem, we propose a delayed ACK, where ACK is not received in the same time slot. The sender receives ACK in the next frame using a piggyback mechanism. The piggyback mechanism saves the communication capacity in networking. When a node must send data and ACK, it sends both of them together in one frame [6]. However, our proposed scheme is different than the existing piggyback mechanism. Because our propose MAC does not have the same destination for data and ACK. The data is sent to the next relaying aircraft, however the ACK is sent to the previous relaying aircraft. Our proposed scheme provides better network utilization because it has one guard time in one time slot. With this increased throughput, we can send about 1.72 times more data than the existing MAC. With this proposed data format, the aircraft sends the weather condition information to the GS using the weather information message format.

This paper is organized as follows: Section 2 briefly reviews the issue of VHF communications and a piggybacking mechanism. Section 3 discusses the proposed protocol and the proposed TDMA MAC structure. The importance of weather information and its message structure is provided in Section 4. Section 5 provides the analysis performance of a proposed scheme. The conclusion is given in Section 6.

## 2. Related Works

Aeronautical communications start analogue voice communication by using DSB-AM (Double-Sideband Amplitude Modulation). Since aeronautical communications need to have a data link, ACARS (Aircraft Communications Addressing and Reporting System) was developed. As it has too low of a data rate, VDL (VHF Data Link) mode 2, 3 and E were developed [1][2][3]. The VDL mode 2 uses a CSMA and the VDL mode 3, E use a TDMA to support VHF aeronautical communications [12] near a Ground Station (GS). For the data and voice communications, the VDL's coverage is up to 480 km.

CSMA-based MAC schemes are not efficient in aeronautical environments because CSMA needs RTS, CTS and ACK messages and they need a long duration guard time. Therefore, many aeronautical systems use TDMA MAC, for example, tactical data link Link-16, Link22 [19]. Another consideration is the underwater environment, which has a long propagation delay like the air environment. Kredo proposed STUMP [20] (Staggered TDMA Underwater MAC Protocol), which utilizes location information to leverage aircraft position diversity and the low propagation speed of the underwater channel.

The ICAO (International Civil Aviation Organization) approved and promoted the transition plan, the CNS/ATM (Communication Navigation Surveillance/Air Traffic Management) system, to support the air traffic expansion and improve aircraft safety. Because of this new plan, in the future, VDL should be able to accommodate increasing data traffic, air-to-air communication, and variable communication modes.

However, VDL mode 3 does not support air-to-air communications, and it only supports voice and data communications between pilots and air traffic controllers. The VDL mode E supports air-to-air communications, but data rate is one-half of VDL mode 3.

It is already known that variable aeronautical communications are more efficient and prompt than fixed VDL communications. For this reason, there are some studies on future aeronautical communication networks.

The VDL VHF communications only supports communication near airports. However, Ho Dac Tu [5] proposed a VHF air-to-air data link for oceanic communications. This new development solves current ATN limitations. First, aircraft could be used as relay systems. Second, it avoids the data traffic bottleneck in the future expansion of aeronautical communications. Third, it supports the new concept of networking in the aeronautical environment, such as NEWSKY (Networking the Sky for Civil aeronautical Communications) project [7], and NOCTARN (New Operational Concept using 3D Adaptable Route Navigation) [8].

The IPv6 based aeronautical communication is bringing large-scale challenges to the new aeronautical communication technologies [17]. In addition, ICAO is working on the standardization of IPv6-based ATN/IPS (Aeronautical Telecommunications Network). For IP based aeronautical communications, Serkan Ayaz [18] proposed an IP-based network and analyzed the handover delay performance.

There are several piggyback mechanisms that help increase the network utilization. TCP ACK is a piggyback mechanism that sends data and ACK to the same nodes [16]. However, in our proposed piggyback mechanism, a node transmits data and ACK to different nodes.

## 3. Proposed TDMA MAC

### A. Long haul aviation communication using TDMA MAC

In this section, we briefly explain Ho Dac Tu's existing long haul aviation communication using TDMA MAC [5]. The ICAO requires all aircraft to periodically send the following information: the aircraft's ID, position information, longitude, latitude, and altitude. Every aircraft is assigned to either generate its own packet or relay a neighbor packet. Ho Dac Tu's proposed scheme uses only one aeronautical VHF channel with an air to air radio relay system based on local mobile ad-hoc networks. To send all of this information, this long haul communication TDMA MAC architecture divides the frame into time slots, where each time slot is a basic unit of access to the network. There are 256 time slots in every frame, and each time slot has a duration of 7.8125 msec. The time slot contains the guard time, data, another guard time, and ACK. The two guard times are dead times in which no data are transmitted; and they occupy 59% of the time slot. Thus, this time slot structure is not efficient. As shown in Figure 1, to solve this problem, we proposed a time slot structure that needs only one guard time.

## B. The proposed protocol and MAC frame

This section explains how each aircraft generates its own packet, relays a neighbor packet, and receives their ACK. We propose a new protocol based on the VHF channel that uses an air to air radio relay system based on local mobile ad-hoc networks. Our proposed protocol is a TDMA that uses contention-free protocol. This TDMA scheme is more effective than CSMA since CSMA's performance is affected by the significant time it takes to send RTS/CTS and ACK data.

Figure 1 shows the proposed frame structure. In this section, time is divided into several frames with a duration of 2 sec. Then, each frame is partitioned into random access and reserved access slots. The frame consists of several time slots, and each time slot is separated by data and guard time. The data part includes 5.515 msec that can contain the aircraft ID and position information. The other part includes a guard time of 2.3 msec to compensate for the free space propagation delay.

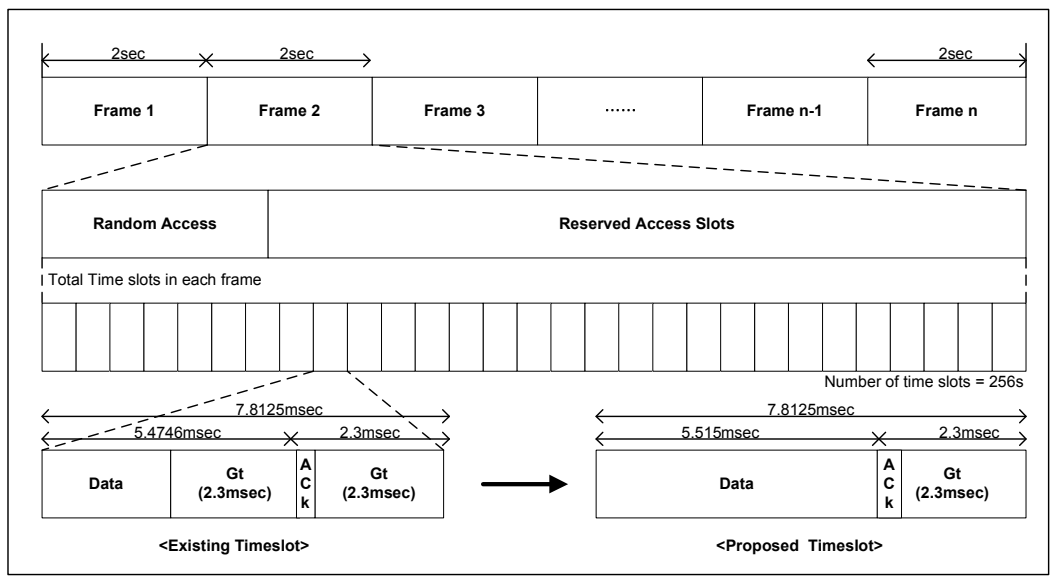

Figure 1. The proposed TDMA MAC frame structure

When an aircraft takes off, it leaves a distance of 540 km with controllers in ground station. If the aircraft wants a reserve slot, it must listen to the network for at least one frame period. During this period, it monitors the random access duration and reserves a time slot. If a time slot is successfully assigned, the aircraft begins transmitting position report data at a predefined interval of several frame periods. Even if successful position report data is received, the receiver does not send ACK or NACK. In the next frame, the receiver relays the received packet to the next relay aircraft using the given by sender time slot. Meanwhile, the relay aircraft sends feedback of ACK or NACK back to the sender via a piggyback mechanism. ACK does not need the destination's address because the sender and receiver both know the ACK's destination. If the sender receives a successful ACK, it will prepare the next transmit position report data at a predefined interval. With a duration of 7.8125 msec per time slot, our proposed scheme transmits data about 2.3 msec faster than the existing scheme depicted in Figure 1. This redundant bit can be used to transmit weather condition information. Weather condition problems (i.e., turbulence, thunderstorms, etc.) have long been a problem in aircraft flight.

## C. Procedure of the proposed protocol

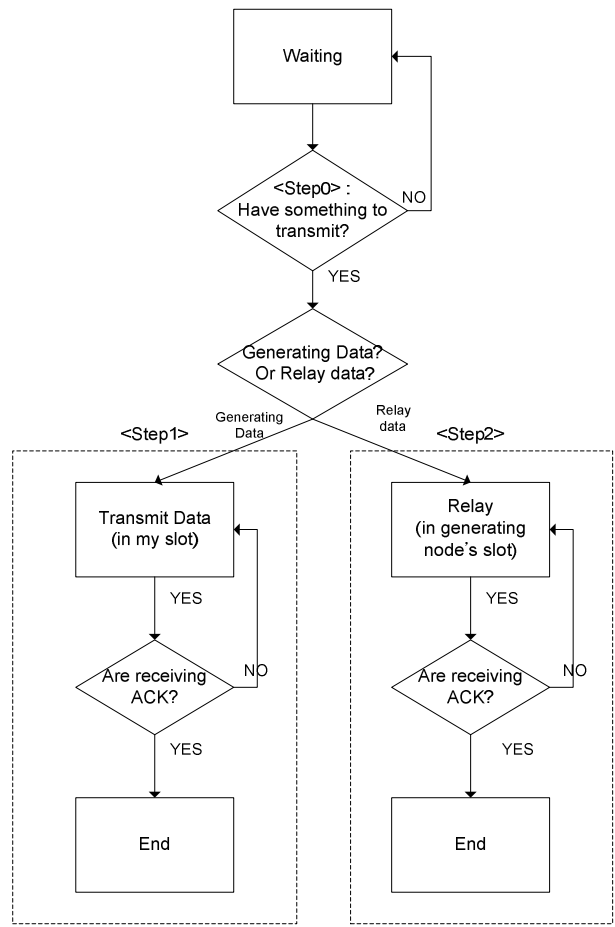

Figure 2. Procedure of the proposed protocol

The proposed protocol procedure is shown in Figure 2, and is explained as follows:

Step 0: The aircraft generates its position report packet at a predefined interval of several frame periods, or it receives a position report packet from a neighboring aircraft.

Step 1: The aircraft transmits its position report data. If the receiver receives a successful position report data, it sends ACK in the same time slot packet in which it arrived, but in the next frame period after the relay data. If the sender does not receive ACK or receives NACK, the sender tries to retransmit its own position report in the next frame.

Step 2: The aircraft receives the position report packet. Then the receiver continues to relay the received packet to the next appropriate relay aircraft in the same time slot packet in which it arrived, but in the next frame period.

### D. A possible case study

Our proposed scheme does not receive ACK in the same time slot. Instead, this mode is a TDD system, which can transmit or receive, but cannot do both at same time. Therefore, when a sender transmits data, it cannot receive data. Thus, we must investigate each of the possible cases.

Figure 3 assumes that all aircraft is adequately separated. It also assumes that each aircraft can generate its own packet at 1 frame, and then periodically generate its own packet and send data at a predefined interval of 3 frame periods. This data transmits to neighboring aircraft; then the receiver relays this data to the next closest neighbor until this data arrives at the GS. Figure 3 also shows how each node how uses their time slot. In this situation, we investigate the two cases where the sender successfully sends data but the receiver does not receive data.

1.) If 'A' successfully sends data to 'B'

Figure 3 shows that, during first frame, 'A' sends the position report data to 'B'. In the second frame, 'B' relays data and sends ACK to 'A', while 'A' listens and waits until it receives ACK. So 'A' is in a listening mode so A can receive ACK.

2.) If 'B' does not receive successful data

We assume that during first frame, 'A' sends the position report data to 'B', but 'B' does not receive the data. Thus, it will not send ACK in the next frame. In the second frame, 'A' waits to receive ACK, but 'A' does not receive ACK. In the third frame, 'A' retransmits the data to 'B'. Thus, this mode has no problem.

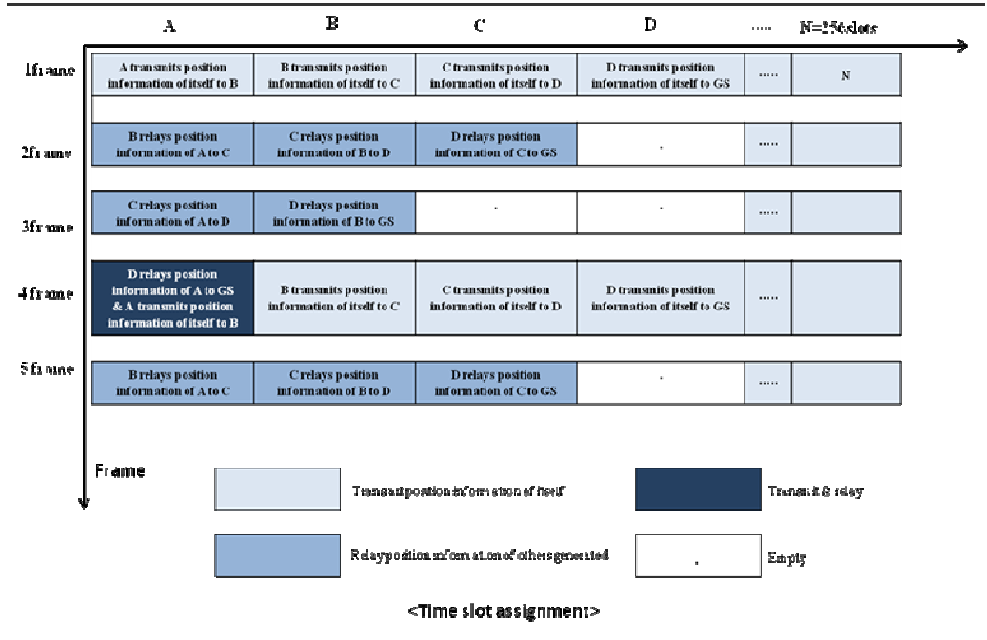

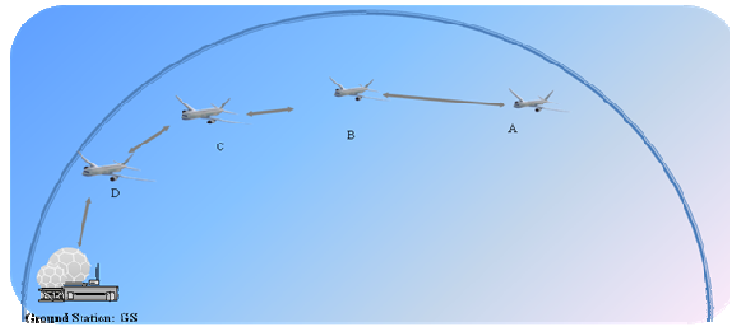

Figure 3. Time slot usage each aviation

## 4. Weather Condition Information

The existing scheme transmits 100 bits in one time slot, but our proposed scheme can transmit 172s bit in one time slot. Additional data can be transmitted periodically. We propose that the weather condition information should be transmitted in the additional time slot. The weather condition information message is in a modified METAR format structure [11].

### A. Why transmit weather condition information?

1.) According to statistics from the National Transportation Safety Board, there were approximately 0.4 accidents per 100,000 departures. And that these accidents, to a large extent, have to do with weather. According to the statistics by a NASA (National Aeronautics and Space Administration) planning group investigating flight accidents caused by the climate, weather contributed to 30% of the accidents [4].

2.) Oceanic weather is challenging because it can be characterized by rapidly changing weather conditions, including heavy rain, severe to extreme turbulence, high winds and gusts, hail, icing, lightning, severe downdrafts, micro bursts, and reduced ceiling and visibility. The domestic weather information system is so successful because it has been researched in numerous studies.

212

But the oceanic weather information system has not been researched very much. Currently, oceanic weather information projects aim to improve these accident statistics by improving weather information available to aviation users. Some examples include AWIN (Aviation Weather Information) and TAMDAR (Troposphere Airborne Meteorological Data Reporting).

3.) Currently, small general aviation aircraft have limited in-flight information on convective weather activity, especially when compared with the information available to larger aircraft [9]. According to statistics, there are approximately 11 weather-related small general aviation aircraft accidents per week, four of which involve fatalities.

For this reason, when aircraft periodically sends weather information to a neighboring aircraft, they can share weather information and avoid areas of bad weather. Currently, NASA is worked on TAMDAR, a data link for collecting weather conditions from aircraft and developing a low-cost sensor [10]. But our proposed scheme does not needs additional equipment and uses a VHF terminal.

## B. Weather information format

We can transmit weather condition information data using the METAR format [11]. METAR format is the most popular format in the world for the transmission of weather data. It is standardized through International Civil Aviation Organization (ICAO).

We can use this format with a few changes to fit our proposed time slot. Figure 4 shows the weather condition information format that fits our proposed scheme. This format uses the following terms:

- Time (hhmmss): The hh is the hour, mm is the minute, and ss is the second.

- Wind (dddss): The value ddd is the wind direction in degrees. The value ss is the wind speed.

- Visibility (ddd_fff): The visibility is where the direction changes.

- Cloud (xhhhy): The x is the cloud level, hhh is the height of the base in 30 m or 100 ft increments.

- Special weather (xxx): Special weather information is listed below in Table 1.

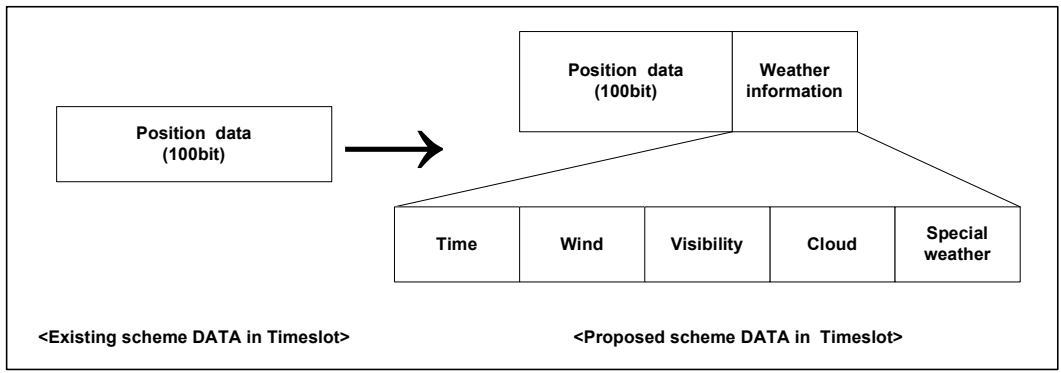

Figure 4. The proposed weather condition information message format

Table 1. Special weather information

| Bit | Special weather |
|---|---|
| 0 | Heavy rain |
| 1 | Severe to extreme turbulence |
| 2 | High winds and gusts |
| 3 | Hail |
| 4 | Icing |
| 5 | Lightning |
| 6 | Severe downdrafts |
| 7 | Microburst |

## 5. Performance Analysis

### A. Link utilization analysis

In this section, we discuss link performance that is the ratio of the total throughput of the network to its data rate, which is defined as [13]

$$U = \frac{Throughput}{data\ rate}$$

The term throughput in this paper is defined the number of bits transmitted per unit time. This is defined as

$$Throughput = \frac{L}{\frac{d}{v} + \frac{L}{R}},$$

where $R$ equals the data rate of the channel; $d$ equals the distance between any two aircraft; $v$ equals the velocity of signal propagation; and $L$ equals the bits contained in the frame.

Figure 5 shows the utilization of network. The vertical axis in the graph indicates the utilization of the network, and the horizontal axis indicates the node's distance. From the result in Figure 5, we can see that the utilization of the network in our proposed TDMA MAC is always higher than the existing TDMA MAC.

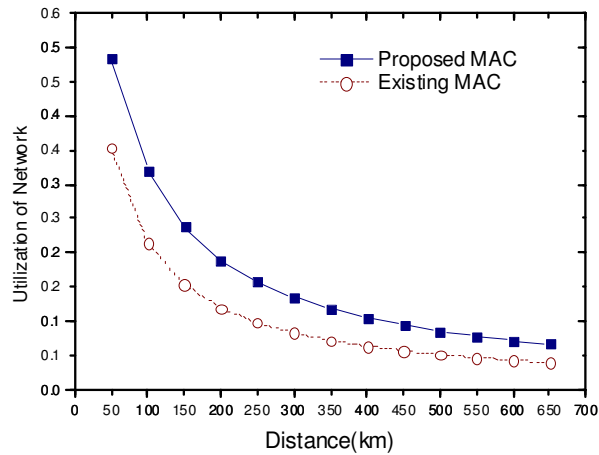

Figure 5. Utilization of the network

**C. Weather information's transmission delay analysis**

Currently, the HF voice is used to send weather information reports, but using the HF voice has its weaknesses. First, there is a potential for human operator error. Second, the HF voice is subject to a number of ionosphere influences which can lead to signal distortion; these influences are dependent upon factors such as the ionosphere layer shape and densities that are functions of geographical and time-varying conditions. Therefore, the HF voice communication success rate averages around 80% [14].

Since many aviation accidents are caused by the weather, weather information is critical. Thus, HF voice communication is not suitable for weather transmission. When a VHF channel with an air-to-air radio relay system is used and based on a mobile ad-hoc network, the packet loss ratio (PLR) is less than 10%. This compares with the normal PLR of 0% [5]. Thus, the VHF data link is more reliable than HF voice communication.

It is also important that the weather conditions of the aircraft's current position are immediately reported to GS. Aviation velocity is 1000 km/h, and aviation's closest approach distance is 90 km. Therefore, the information from the closest approach will take 5.4 minutes to arrive. However, if we consider the pilot's recognition ability, determination, and aviation angular velocity, the weather information transmission can be shorter than 5.4 min. Therefore, we must consider the weather information delay of arrival to GS. In this paper, the total time of transmitting weather conditions to GS is defined as the "weather notification delay." The weather notification delay is 1-2 minutes in HF communication [14]. Our proposed scheme's weather notification delay $d_t$ can be obtained by the equation:

$$d_t = N_i \times H_i \times T_f,$$

where $N_i$ denotes the average number of transmissions for each link, $H_i$ denotes the number of relaying nodes from the source to the destination, and $T_f$ is the frame duration. The variable

215

$N_i$ is characterized by the expected value of a geometrically distributed packet loss ratio $\rho \{ 0 \leq \rho \leq 0.1 \}$, where $N_i$ is calculated as follows:

$$N_i = \frac{1}{1-\rho}$$

The variable $H_i$ is characterized by the interval of nodes and the total distance. $H_i$ is calculated as follows:

$$H_i = \frac{d_{gs}}{d_{int}},$$

where $d_{gs}$ is the distance to the ground station, and $d_{int}$ is the interval of nodes. On the oceanic areas that are out of the radar's range, the safety interval is required to be much longer at 90 km.

If $d_{int}$ is 90 km, then $H_i$ has the maximum relay nodes. We denoted $H_{i\_max}$. If $d_{int}$ is 678 km (distance of line of sight equals 678 km) [5], then $H_i$ has the minimum relay nodes. We denoted $H_{i\_min}$. $T_f$ is the frame duration. Our proposed frame duration is 2 second. In this paper, we assume the distance between the GS and the aircraft is 4900 km (Pacific's East West radius). We also assume that the aircraft sends the weather condition information to the GS by relaying the information to other aircraft.

Figure 6 compares the weather information transmission delay for HF voice communication to our proposed schemes when $\rho=0$ and $\rho=0.1$. Weather notification delay is average delay in case $H_i$ has maximum relay nodes and minimum relay nodes. In our proposed scheme, $\rho=0.1$ means that some aircraft didn't find any route to relay the packet because the aircraft traffic is more sparse at night. The weather notification delay of $\rho=0$ means that the aircraft were able to find a route to relay the packet, which occurred most of time. The weather notification delay of the HF voice communication is not related to distance. Therefore, we used an average of the HF voice communication's weather notification delay.

We can see the weather notification delay increase almost linearly with the distance to the GS. In addition, the weather notification delay is more effective than the HF voice communication because the VHF communication avoids the HF communication's weaknesses, and our proposed scheme uses data communication. Also, $\rho=0.1$ has more delay than $\rho=0$ because retransmission is more frequent.

Figure 7 shows that the weather notification delay is more effective when there is a greater distance between aircraft. When the aircraft is out of the airport radar's range, the aircraft's safety interval must be much longer at 90 km and the data propagation distance is about 780 km. Therefore, the horizontal axis indicates the distance between the two aircraft is 90 km to 790 km. Figure 7 illustrate that the aircraft's long distance has a shorter delay. When the aircraft's distance is 90 km, relaying occurs frequently; when the aircraft's distance is 780 km, relaying occurs sparsely.

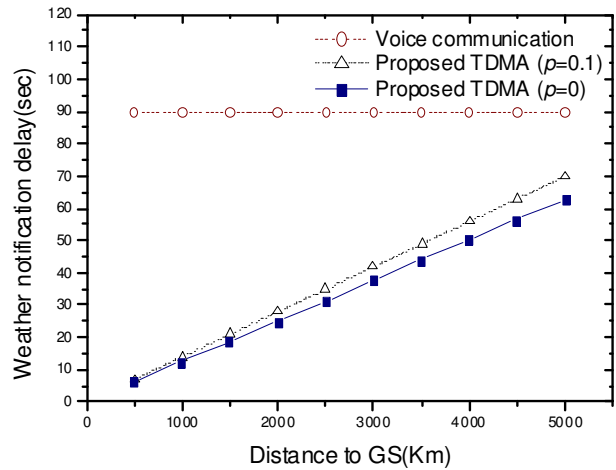

Figure 6. Weather notification delay of HF voice and the proposed scheme with distance to GS

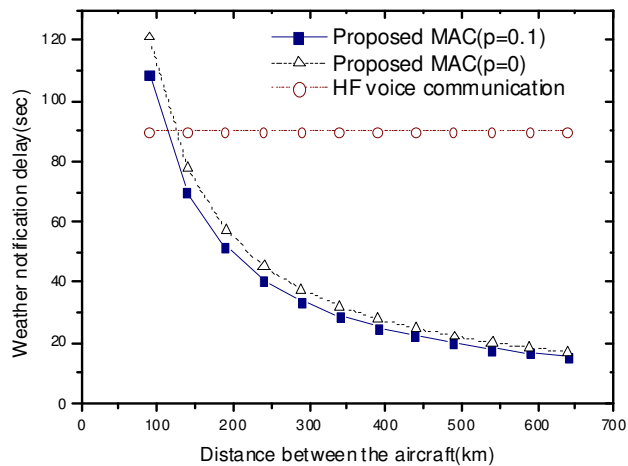

Figure 7. Weather notification delay of the proposed scheme and HF voice communication with distance between the aircraft

C. End to end throughput

In the previous formulation we calculate the end to end delay time. Here, we calculate the end to end throughput using the end to end delay time. The end to end Throughput $S$ can be obtained by the equation:

$$S = \frac{T_d \times H_{i\_average}}{d_t},$$

217

where $T_d$ denotes the number of transmitted bits in one slot, $H_{i\_average}$ denotes average of the number of aircraft which is defined in section 4.C, $d_t$ denotes the average end to end delay which is defined in section 4.C. In this equation, we can see end to end throughput using the number of bits that arrive to GS. From the result in Figure 8, the end to end throughput in the proposed TDMA MAC is always higher than the existing TDMA MAC. As we can see, the end to end throughput of proposed MAC (p=0.1) is higher than the existing MAC (p=0) even in the case that PLR is 10%.

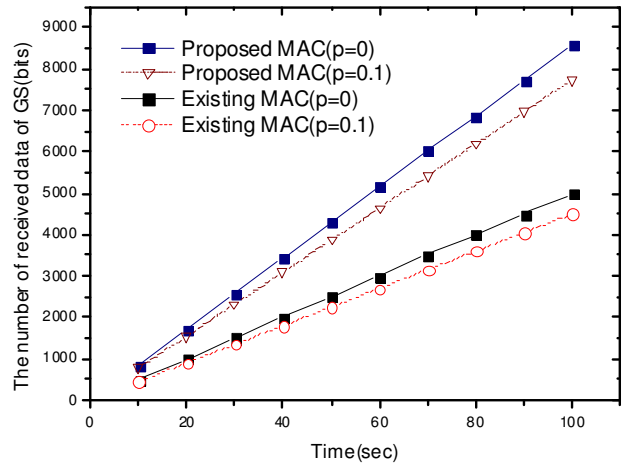

Figure 8. The end to end throughput of proposed MAC and Existing MAC with distance to GS

## 6. Conclusion

In this paper, we proposed a piggybacking mechanism ACK. In our scheme, the aircraft sends position reporting data and the receiver receives successful position reporting data; the receiver does not send ACK or NACK. In the next frame, the receiver continues to relay the received packet to the next appropriate relay aircraft; this is sent via the same time slot packet in which it arrived. In addition, the relay aircraft will send feedback of ACK or NACK to the sender through a piggyback mechanism. Therefore, our proposed MAC needs one guard time in each time slot. Through a performance analysis, we were able to verify the advantage of using our proposed MAC in terms of link utility and network end to end throughput. With this increased throughput, we can send about 1.72 times more data than the existing MAC. With this proposed data format, the aircraft sends the weather condition information to the GS using the weather information message format. We also showed that weather condition notification using our scheme achieves a much lower transmission delay than the HF voice communication.

### ACKNOWLEDGEMENTS

"This research was supported by the MKE(The Ministry of Knowledge Economy), Korea, under the ITRC(Information Technology Research Center)support program supervised by the NIPA(National IT Industry Promotion Agency" (NIPA-2011-(C1090-1121-0011))

**Authors**

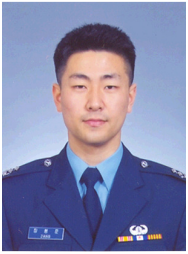
Hyungjun Jang was born in Cheju, South Korea, in 1980. He received B.S in International Relationship in 2004 at Korea Air Force Academy (KAFA). Since Marh 2008, he studied for M.S & ph.D degree in Network Centric Warfare of Graduate School of Ajou University. His research interests are related to tactical data link, civil aeronautical communication and air traffic control.

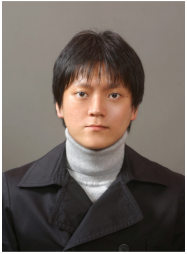
Hongjun Noh was born in Seoul, South Korea, in 1982. He graduated from Ajou University of KOREA, March 2008. From March 2008 he has been Master&Ph.D candidate at Computer Engineering of Graduate School of Ajou University. His research interests are related to tactical data link and Aeronautical Communication.

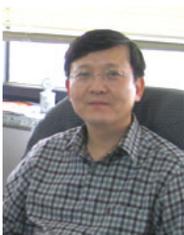
Jaesung Lim received the B.S. in electronic engineering from Ajou University, Korea, in 1983, and the M.S. and Ph.D, in electrical engineering from Korea Advanced Institute of Science and Technology (KAIST), in 1985 and 1994, respectively. In 1985, he started as a reseacher at the DAEWOO telecommunication. In April 1988, he joined the institute of DigiCom, and was engaged in research and development of data modem, radar signal processing and packet data systems. From 1995 to 1997, he served as a senior engineer in the Central Research & Development Center of SK Telecom, where he did research on wireless data communications for cellular and paging networks. Since March 1998 he has been with Ajou University, where he is a professor of the Graduate School of Information and Communication Technology, teaching and doing research in the areas of wireless, mobile, and tactical communications and networks. He has also been Director of AJOU-TNRC(Tactical Networks Research Center) since 2006.